\documentclass[prl,twocolumn,amsmath,amssymb]{revtex4}
\usepackage{dsfont}
\usepackage{graphicx}
\usepackage{pstricks}

\newcommand{\ket}[1]{|#1 \rangle}

\newcommand{\R}{\text{R}}

\begin{document}

\title{On the Optimality of Basis Transformations \\
to Secure Entanglement Swapping Based QKD Protocols}

\author{Stefan Schauer, Martin Suda}
\affiliation{AIT Austrian Institute of Technology GmbH, Donau-City-Str. 1, A-1220 Vienna, Austria}

\begin{abstract}
In this article, we discuss the optimality of basis transformations as a security measure for quantum key distribution protocols based on entanglement swapping. To estimate the security, we focus on the information an adversary obtains on the raw key bits from a generic version of a collective attack strategy. In the scenario described in this article, the application of general basis transformations serving as a counter measure by one or both legitimate parties is analyzed. In this context, we show that the angles, which describe these basis transformations can be optimized compared to the application of a Hadamard operation, which is the standard basis transformation recurrently found in literature. As a main result, we show that the adversary's information can be reduced to an amount of $I_{AE} \simeq 0.20752$ when using a single basis transformation and to an amount of $I_{AE} \simeq 0.0548$ when combining two different basis transformations. This is less than half the information compared to other protocols using a Hadamard operation and thus represents an advantage regarding the security of entanglement swapping based protocols. 
\end{abstract}

\maketitle

\section{Introduction}

Quantum key distribution (QKD) is an important application of quantum mechanics and QKD protocols have been studied at length in theory and in practical implementations \cite{BB84,Eke91,BBM92,Bru98,MZG96,PF04,PPM08,PPA09}. Most of these protocols focus on prepare and measure schemes, where single qubits are in transit between the communication parties Alice and Bob. The security of these protocols has been discussed in depth and security proofs have been given for example in \cite{Lut96,Lut00,SP00}. In addition to these prepare and measure protocols several protocols based on the phenomenon of entanglement swapping have been introduced \cite{Cab00a,Cab01,Cab00b,Son04,LWWSZ06}. In these protocols, entanglement swapping is used to obtain correlated measurement results between the legitimate communication parties, Alice and Bob. In other words, each party performs a Bell state measurement and due to entanglement swapping their results are correlated and further on used to establish a secret key. 
\par
\begin{figure}
\centering
\psset{unit=0.6cm}
\tiny
\scalebox{1} 
{
\begin{pspicture}(0,-3.0)(14.245,3.5)
\psframe[linewidth=0.02,dimen=outer,fillstyle=solid,fillcolor=lightgray](2.0,3.0)(0.0,1.0)
\psframe[linewidth=0.02,dimen=outer,fillstyle=solid,fillcolor=lightgray](5.0,3.0)(3.0,1.0)
\psline[linewidth=0.02cm,fillcolor=black,dotsize=0.07055555cm 2.0]{*-*}(1.0,2.5)(1.0,1.5)
\psline[linewidth=0.02cm,fillcolor=black,dotsize=0.07055555cm 2.0]{*-*}(4.0,2.5)(4.0,1.5)
\rput(1.0,3.25){Alice}
\rput(4.0,3.25){Bob}
\rput(4.0,1.25){$\ket{\Phi^+}$}
\rput(1.0,1.25){$\ket{\Phi^+}$}
\rput(0.7,2.5){\tiny $1$}
\rput(0.7,1.5){\tiny $2$}
\rput(4.3,2.5){\tiny $3$}
\rput(4.3,1.5){\tiny $4$}
\rput(6.0,2.0){(1)}
\psframe[linewidth=0.02,dimen=outer,fillstyle=solid,fillcolor=lightgray](10.0,3.0)(8.0,1.0)
\psframe[linewidth=0.02,dimen=outer,fillstyle=solid,fillcolor=lightgray](13.0,3.0)(11.0,1.0)
\psline[linewidth=0.02cm,fillcolor=black,dotsize=0.07055555cm 2.0]{*-*}(9.0,2.5)(12.0,2.5)
\psline[linewidth=0.02cm,fillcolor=black,dotsize=0.07055555cm 2.0]{*-*}(12.0,1.5)(9.0,1.5)
\pscircle[linewidth=0.02,linestyle=dashed,dash=0.1cm 0.1cm,dimen=outer](9.0,2.5){0.2}
\rput(9.0,3.25){Alice}
\rput(12.0,3.25){Bob}
\rput(8.5,2.5){$T_x$}
\rput(14.0,2.0){(2)}
\psframe[linewidth=0.02,dimen=outer,fillstyle=solid,fillcolor=lightgray](2.0,-1.0)(0.0,-3.0)
\psframe[linewidth=0.02,dimen=outer,fillstyle=solid,fillcolor=lightgray](5.0,-1.0)(3.0,-3.0)
\psframe[linewidth=0.02,linestyle=dashed,dash=0.1cm 0.1cm,framearc=1.0,dimen=outer](1.2,-1.3)(0.8,-2.7)
\psline[linewidth=0.02cm,fillcolor=black,dotsize=0.07055555cm 2.0]{*-*}(1.0,-1.5)(4.0,-1.5)
\psline[linewidth=0.02cm,fillcolor=black,dotsize=0.07055555cm 2.0]{*-*}(4.0,-2.5)(1.0,-2.5)
\pscircle[linewidth=0.02,linestyle=dashed,dash=0.1cm 0.1cm,dimen=outer](4.0,-1.5){0.2}
\rput(1.0,-0.75){Alice}
\rput(4.0,-0.75){Bob}
\rput(4.5,-1.5){$T_x$}
\rput(6.0,-2.0){(3)}
\psframe[linewidth=0.02,dimen=outer,fillstyle=solid,fillcolor=lightgray](10.0,-1.0)(8.0,-3.0)
\psframe[linewidth=0.02,dimen=outer,fillstyle=solid,fillcolor=lightgray](13.0,-1.0)(11.0,-3.0)
\psline[linewidth=0.02cm,fillcolor=black,dotsize=0.07055555cm 2.0]{*-*}(9.0,-1.5)(9.0,-2.5)
\psline[linewidth=0.02cm,fillcolor=black,dotsize=0.07055555cm 2.0]{*-*}(12.0,-2.5)(12.0,-1.5)
\rput(9.0,-0.75){Alice}
\rput(12.0,-0.75){Bob}
\rput(8.7,-2.5){\tiny $3$}
\rput(8.7,-1.5){\tiny $1$}
\rput(12.3,-2.5){\tiny $4$}
\rput(12.3,-1.5){\tiny $2$}
\rput(9.0,-2.75){$\ket{\Psi^+}$}
\rput(12.0,-2.75){$\ket{\Psi^+}$}
\rput(14.0,-2.0){(4)}
\end{pspicture} 
}
\caption{Sketch of a standard setup for an entanglement swapping based QKD protocol using a basis transformation $T_x$.}
\label{fig:ES-QKD}
\end{figure}
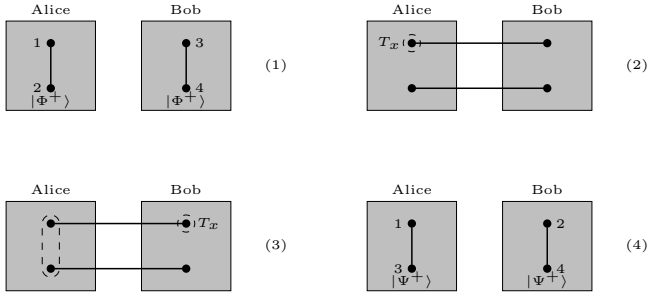

Entanglement swapping has been introduced by Bennett et al. \cite{BBCJ93}, Zukowski et al. \cite{ZZHE93} as well as Yurke and Stolen \cite{YS92}, respectively. It provides the unique possibility to generate entanglement from particles that never interacted in the past. In detail, Alice and Bob share two Bell states of the form $\ket{\Phi^+}_{12}$ and $\ket{\Phi^+}_{34}$ such that afterwards Alice is in possession of qubits 1 and 3 and Bob of qubits 2 and 4 (cf. Figure \ref{fig:ES-QKD}). Then Alice performs a complete Bell state measurement on the two qubits in her possession, which results in
\begin{equation} 
\begin{aligned}
\ket{\Phi^+}_{12} \otimes \ket{\Phi^+}_{34} = \frac{1}{2} \Bigl(
 &\ket{\Phi^+}  \ket{\Phi^+} + \ket{\Phi^-}  \ket{\Phi^-} \\
+&\ket{\Psi^+}  \ket{\Psi^+} + \ket{\Psi^-}  \ket{\Psi^-} \Bigr)_{1324}
\end{aligned}
\label{eq:EntSwapping}
\end{equation}
After the measurement, the qubits 2 and 4 at Bob's side collapse into a Bell state although they originated at completely different sources. Moreover, the state of Bob's qubits depends on Alice's measurement result. As presented in eq. (\ref{eq:EntSwapping}), Bob always obtains the same result as Alice when performing a Bell state measurement on his qubits.
\par
The security of QKD protocols based on entanglement swapping has been discussed on the surface so far. It has only been shown that these protocols are secure against intercept-resend attacks and basic collective attacks (cf. for example \cite{Cab00a,Cab01,Son04}). Therefore, we analyze a general version of a collective attack where the adversary tries to simulate the correlations between Alice and Bob \cite{SS08}. A basic technique to secure a QKD protocol is to use a basis transformation, usually a Hadamard operation, to make it easier to detect an adversary as implemented, for example, in the prepare and measure schemes described in \cite{BB84} and \cite{BBM92}. Hence, we analyze the security with respect to a general basis transformation defined by the angles $\theta_A$ and $\phi_A$ applied by Alice and a transformation defined by the angles $\theta_B$ and $\phi_B$ applied by Bob. In the course of that, we are going to identify which values for $\theta_A$, $\phi_A$, $\theta_B$ and $\phi_B$ are optimal such that an adversary has only a minimum amount of information on the secret raw key.
\par
In the following section, we give a short review of the \emph{simulation attack}, a generic collective attack strategy where an adversary applies a six-qubit state to eavesdrop Bob's measurement result. A detailed discussion of this attack strategy can be found in \cite{SS08}. Further, we look at the general definition of basis transformations and their effect onto Bell states and entanglement swapping. Using the definitions of general basis transformations, we discuss in the next sections the effects on the security of entanglement swapping based QKD protocols regarding two scenarios: we distinguish between the application of a basis transformation by only one communication party and the combined application of two different basis transformations by each of the communication parties. Additionally, we analyze for each of these scenarios the application of a simple transformation with only one degree of freedom and a general transformation. In the end, we sum up the implications of the results on the security of entanglement based QKD protocols.

\section{The Simulation Attack Strategy} \label{sec:SimAttack}
In entanglement swapping based QKD protocols like \cite{Cab00a,Cab01,Cab00b,Son04,LWWSZ06} Alice and Bob rest their security check onto the correlations between their respective measurement results coming from the entanglement swapping (cf. eq. (\ref{eq:EntSwapping})). If these correlations are violated to a certain amount, Alice and Bob have to assume that an eavesdropper is present. Hence, a general version of a collective attack has the following basic idea: the adversary Eve tries to find a multi-qubit state, which preserves the correlation between the two legitimate parties. Further, she introduces additional qubits to distinguish between Alice's and Bob's respective measurement results. If she is able to find such a state, Eve stays undetected during her intervention and is able to obtain a certain amount of information about the key. In a previous article \cite{SS08}, we already described such a collective attack called \textit{simulation attack} for a specific protocol \cite{LWWSZ06}. The generalization is straight forward as described in the following paragraphs.
\par 
It has been pointed out in detail in \cite{SS08} that Eve uses four qubits to simulate the correlations between Alice and Bob and she further introduces additional systems, i.e. $\ket{\varphi_i}$, to distinguish between Alice's different measurement results. This leads to the state
\begin{equation}
\begin{aligned}
\ket{\delta} = \frac{1}{2} \Bigl(
&\ket{\Phi^+}\ket{\Phi^+}\ket{\varphi_1} + \ket{\Phi^-}\ket{\Phi^-}\ket{\varphi_2} \\
&\ket{\Psi^+}\ket{\Psi^+}\ket{\varphi_3} + \ket{\Psi^-}\ket{\Psi^-}\ket{\varphi_4} 
\Bigr)_{PRQSTU} 
\end{aligned}
\label{eq:DeltaCorrelations}
\end{equation}
which is a more general version than described in \cite{SS08}. This state preserves the correlation of Alice's and Bob's measurement results coming from the entanglement swapping (cf. eq. (\ref{eq:EntSwapping})): from eq. (\ref{eq:DeltaCorrelations}) it is easy to see that Alice obtains one of the four Bell states when performing a Bell state measurement on qubits $P$ and $R$. This measurement leaves Bob's qubits $Q$ and $S$ in a Bell state fully correlated to Alice's result. Accordingly, Eve's qubits $T$ and $U$ are in one of the auxiliary states $\ket{\varphi_i}$ she prepared. 
\par
Eve has to choose the auxiliary systems $\ket{\varphi_i}$ such that
\begin{equation}
\langle \varphi_i | \varphi_j \rangle = 0 \qquad i,j \in \{1, ... ,4\} \;\; i \neq j
\label{eq:Varphi}
\end{equation} 
which allows her to perfectly distinguish between Alice's and Bob's respective measurement results. Thus, she is able to eavesdrop Alice's and Bob's measurement results and obtains full information about the classical raw key generated out of them.
\par
There are different ways for Eve to distribute the state $\ket{\delta}_{P-U}$ between Alice and Bob. One possibility is that Eve is in possession of Alice's and Bob's source and generates $\ket{\delta}_{P-U}$ instead of Bell states. This is a rather strong assumption, because the sources are usually located at Alice's or Bob's laboratory, which should be a secure environment. Eve's second possibility is to intercept the qubits 2 and 3 flying from Alice to Bob and vice versa and to use entanglement swapping to distribute the state $\ket{\delta}$. This is a straight forward method as already described in \cite{SS08}.
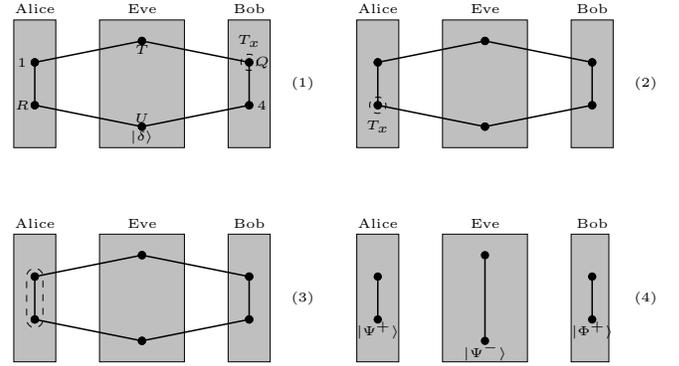
\begin{figure}
\centering
\psset{unit=0.57cm}
\tiny
\scalebox{1} 
{
\begin{pspicture}(0,-4.0)(15.045,4.5)
\psframe[linewidth=0.02,dimen=outer,fillstyle=solid,fillcolor=lightgray](1.0,4.0)(0.0,1.0)
\psframe[linewidth=0.02,dimen=outer,fillstyle=solid,fillcolor=lightgray](6.0,4.0)(5.0,1.0)
\psframe[linewidth=0.02,dimen=outer,fillstyle=solid,fillcolor=lightgray](4.0,4.0)(2.0,1.0)
\psline[linewidth=0.02cm,fillcolor=black,dotsize=0.07055555cm 2.0]{*-*}(3.0,3.5)(0.5,3.0)
\psline[linewidth=0.02cm,fillcolor=black,dotsize=0.07055555cm 2.0]{*-*}(0.5,3.0)(0.5,2.0)
\psline[linewidth=0.02cm,fillcolor=black,dotsize=0.07055555cm 2.0]{*-*}(0.5,2.0)(3.0,1.5)
\psline[linewidth=0.02cm,fillcolor=black,dotsize=0.07055555cm 2.0]{*-*}(3.0,1.5)(5.5,2.0)
\psline[linewidth=0.02cm,fillcolor=black,dotsize=0.07055555cm 2.0]{*-*}(5.5,2.0)(5.5,3.0)
\psline[linewidth=0.02cm,fillcolor=black,dotsize=0.07055555cm 2.0]{*-*}(5.5,3.0)(3.0,3.5)
\pscircle[linewidth=0.02,linestyle=dashed,dash=0.1cm 0.1cm,dimen=outer](5.5,3.0){0.2}
\rput(0.5,4.25){Alice}
\rput(3.0,4.25){Eve}
\rput(5.5,4.25){Bob}
\rput(5.5,3.5){$T_x$}
\rput(3.0,1.25){$\ket{\delta}$}
\rput(3.0,3.3){\tiny $T$}
\rput(0.2,3.0){\tiny $1$}
\rput(5.8,3.0){\tiny $Q$}
\rput(0.2,2.0){\tiny $R$}
\rput(5.8,2.0){\tiny $4$}
\rput(3.0,1.7){\tiny $U$}
\rput(6.75,2.5){(1)}
\psframe[linewidth=0.02,dimen=outer,fillstyle=solid,fillcolor=lightgray](9.0,4.0)(8.0,1.0)
\psframe[linewidth=0.02,dimen=outer,fillstyle=solid,fillcolor=lightgray](14.0,4.0)(13.0,1.0)
\psframe[linewidth=0.02,dimen=outer,fillstyle=solid,fillcolor=lightgray](12.0,4.0)(10.0,1.0)
\psline[linewidth=0.02cm,fillcolor=black,dotsize=0.07055555cm 2.0]{*-*}(8.5,2.0)(11.0,1.5)
\psline[linewidth=0.02cm,fillcolor=black,dotsize=0.07055555cm 2.0]{*-*}(11.0,1.5)(13.5,2.0)
\psline[linewidth=0.02cm,fillcolor=black,dotsize=0.07055555cm 2.0]{*-*}(8.5,3.0)(8.5,2.0)
\psline[linewidth=0.02cm,fillcolor=black,dotsize=0.07055555cm 2.0]{*-*}(13.5,2.0)(13.5,3.0)
\psline[linewidth=0.02cm,fillcolor=black,dotsize=0.07055555cm 2.0]{*-*}(11.0,3.5)(8.5,3.0)
\psline[linewidth=0.02cm,fillcolor=black,dotsize=0.07055555cm 2.0]{*-*}(11.0,3.5)(13.5,3.0)
\pscircle[linewidth=0.02,linestyle=dashed,dash=0.1cm 0.1cm,dimen=outer](8.5,2.0){0.2}
\rput(8.5,4.25){Alice}
\rput(11.0,4.25){Eve}
\rput(13.5,4.25){Bob}
\rput(8.5,1.5){$T_x$}
\rput(14.75,2.5){(2)}
\psframe[linewidth=0.02,dimen=outer,fillstyle=solid,fillcolor=lightgray](1.0,-1.0)(0.0,-4.0)
\psframe[linewidth=0.02,dimen=outer,fillstyle=solid,fillcolor=lightgray](6.0,-1.0)(5.0,-4.0)
\psframe[linewidth=0.02,dimen=outer,fillstyle=solid,fillcolor=lightgray](4.0,-1.0)(2.0,-4.0)
\psframe[linewidth=0.02,linestyle=dashed,dash=0.1cm 0.1cm,framearc=1.0,dimen=outer](0.7,-1.8)(0.3,-3.2)
\psline[linewidth=0.02cm,fillcolor=black,dotsize=0.07055555cm 2.0]{*-*}(5.5,-3.0)(5.5,-2.0)
\psline[linewidth=0.02cm,fillcolor=black,dotsize=0.07055555cm 2.0]{*-*}(0.5,-3.0)(3.0,-3.5)
\psline[linewidth=0.02cm,fillcolor=black,dotsize=0.07055555cm 2.0]{*-*}(0.5,-2.0)(0.5,-3.0)
\psline[linewidth=0.02cm,fillcolor=black,dotsize=0.07055555cm 2.0]{*-*}(3.0,-3.5)(5.5,-3.0)
\psline[linewidth=0.02cm,fillcolor=black,dotsize=0.07055555cm 2.0]{*-*}(3.0,-1.5)(0.5,-2.0)
\psline[linewidth=0.02cm,fillcolor=black,dotsize=0.07055555cm 2.0]{*-*}(3.0,-1.5)(5.5,-2.0)
\rput(0.5,-0.75){Alice}
\rput(3.0,-0.75){Eve}
\rput(5.5,-0.75){Bob}
\rput(6.75,-2.5){(3)}
\psframe[linewidth=0.02,dimen=outer,fillstyle=solid,fillcolor=lightgray](9.0,-1.0)(8.0,-4.0)
\psframe[linewidth=0.02,dimen=outer,fillstyle=solid,fillcolor=lightgray](13.9,-1.0)(13.0,-4.0)
\psframe[linewidth=0.02,dimen=outer,fillstyle=solid,fillcolor=lightgray](12.0,-1.0)(10.0,-4.0)
\psline[linewidth=0.02cm,fillcolor=black,dotsize=0.07055555cm 2.0]{*-*}(8.5,-2.0)(8.5,-3.0)
\psline[linewidth=0.02cm,fillcolor=black,dotsize=0.07055555cm 2.0]{*-*}(11.0,-1.5)(11.0,-3.5)
\psline[linewidth=0.02cm,fillcolor=black,dotsize=0.07055555cm 2.0]{*-*}(13.5,-2.0)(13.5,-3.0)
\rput(8.5,-0.75){Alice}
\rput(11.0,-0.75){Eve}
\rput(13.5,-0.75){Bob}
\rput(8.5,-3.25){$\ket{\Psi^+}$}
\rput(11.0,-3.75){$\ket{\Psi^-}$}
\rput(13.5,-3.25){$\ket{\Phi^+}$}
\rput(14.75,-2.5){(4)}
\end{pspicture} 
}
\caption{Illustration of the simulation attack on a standard setup for an entanglement swapping based QKD protocol using a basis transformation $T_x$.}
\label{fig:SaonES-QKD}
\end{figure} 
\par
In detail, Eve distributes qubits $P$, $Q$, $R$ and $S$ between Alice and Bob using entanglement swapping such that Alice is in possession of qubits $1$ and $R$ and Bob is in possession of qubits $Q$ and $4$ (cf. (1) in Figure \ref{fig:SaonES-QKD}). When Alice performs a Bell state measurement on qubits $1$ and $R$ the state of qubits $Q$ and $4$ collapses into the same Bell state, which Alice obtained from her measurement (cf. eq. (\ref{eq:DeltaCorrelations}) and picture (4) in Figure \ref{fig:SaonES-QKD}). In particular, if Alice obtains $\ket{\Phi^+}_{1R}$ the state of the remaining qubits is
\begin{equation}
\ket{\Phi^+}_{Q4}\ket{\varphi_1}_{TU}
\end{equation} 
and similarly for Alice's other results $\ket{\Phi^-}$ and $\ket{\Psi^{\pm}}$. This is the exact correlation Alice and Bob would expect from entanglement swapping if no adversary is present (cf. eq. (\ref{eq:EntSwapping}) from above). Hence, Eve stays undetected when Alice and Bob compare some of their results in public to check for eavesdroppers. The auxiliary system $\ket{\varphi_i}$ remains at Eve's side and its state is completely determined by Alice's measurement result. Therefore, Eve has full information on Alice's and Bob's measurement results and is able to perfectly eavesdrop the classical raw key. 
\par
We want to stress that the state $\ket{\delta}$ is generic for all protocols where 2 qubits are exchanged between Alice and Bob during one round of key generation as, for example, the QKD protocols presented by Song \cite{Son04}, Li et al. \cite{LWWSZ06} or Cabello \cite{Cab00a}. As already pointed out in \cite{SS08}, the state $\ket{\delta}$ can also be used for different initial Bell states. For protocols with a higher number of qubits the state $\ket{\delta}$ has to be extended accordingly.  

\section{Basis Transformations} \label{sec:BT}
In QKD, the most common way to detect the presence of an adversary is to use a random application of a basis transformation by one of the legitimate communication parties. This method can be found in prepare an measure protocols (e.g. in \cite{BB84} or \cite{BBM92}) as well as entanglement swapping based protocols (e.g. in \cite{Cab01,Son04} or the improved version of the protocol in \cite{LWWSZ06}). The idea is to randomly alter the initial state to make it impossible for an adversary to eavesdrop the information transmitted without introducing a certain error rate, i.e. without being detected. The operation most commonly used in these protocols is the Hadamard operation described by the matrix 
\begin{equation}
\frac{1}{\sqrt{2}} \begin{pmatrix}1 & 1 \\ 1 & -1 \end{pmatrix}
\label{eq:OpHadamard}
\end{equation}
which is a transformation from the $Z$- into the $X$-basis. For our further discussion, the $X$-basis is given by the states $\ket{x+}$ and $\ket{x-}$ with 
\begin{equation}
\ket{x\pm} = \frac{1}{\sqrt{2}} \Bigl( \ket{0} \pm \ket{1} \Bigr)
\label{eq:XBasis}
\end{equation}
and the $Z$-basis is the computational basis consisting of the states $\ket{0}$ and $\ket{1}$. The effect of the Hadamard operation is simply described as
\begin{equation}
\ket{0} \longmapsto \ket{x+} \quad \ket{1} \longmapsto \ket{x-}.
\end{equation}
In general, a transformation $T_x$ from the $Z$ basis into the $X$-basis can be described as a rotation about the $X$-axis by some angle $\theta$ combined with two rotations about the $Z$-axis by some angle $\phi$, i.e.
\begin{equation}
T_x\bigl(\theta,\phi\bigr) = e^{i\phi} R_z\bigl(\phi\bigr) R_x\bigl(\theta\bigr) R_z\bigl(\phi\bigr).
\label{eq:OpTransX} 
\end{equation}
The rotations about the $X$- or $Z$-axis are described in the most general way by the operators (cf. for example \cite{NC00} for further details on rotation operators)
\begin{equation}
\begin{aligned}
\R_x\bigl(\theta\bigr) &= \begin{pmatrix}\cos \frac{\theta}{2} & -i\sin \frac{\theta}{2} \\ -i\sin \frac{\theta}{2} & \cos \frac{\theta}{2} \end{pmatrix} \\
\R_z\bigl(\theta\bigr) &= \begin{pmatrix}e^{-i \theta /2} & 0 \\ 0 & e^{i \theta /2} \end{pmatrix}.
\end{aligned}
\label{eq:OpRotation}
\end{equation}
For the sake of completeness we want to add that a rotation about the $Y$-axis is similarly defined by the operator
\begin{equation}
\R_y\bigl(\theta\bigr) = \begin{pmatrix}\cos \frac{\theta}{2} & -\sin \frac{\theta}{2} \\ \sin \frac{\theta}{2} & \cos \frac{\theta}{2} \end{pmatrix}
\end{equation}
but will not be used in the context of this article. Based on these operators, we directly obtain the matrix representation for $T_x(\theta,\phi)$ as
\begin{equation}
\begin{aligned}
T_x\bigl(\theta,\phi\bigr) &= \begin{pmatrix}
              \cos \frac{\theta}{2} & -i\;e^{i\phi} \sin \frac{\theta}{2} \\ 
-i\;e^{i\phi} \sin \frac{\theta}{2} &    e^{2i\phi}\cos \frac{\theta}{2} \end{pmatrix} \\
\end{aligned}
\label{eq:OpTransXMatrix} 
\end{equation}
and the effect of $T_x(\theta,\phi)$ on the computational basis
\begin{equation}
\begin{aligned}
T_x\bigl(\theta,\phi\bigr) \ket{0} 
&= \cos \frac{\theta}{2} \ket{0} - i\;e^{i\phi} \sin \frac{\theta}{2} \ket{1} \\
T_x\bigl(\theta,\phi\bigr) \ket{1} 
&=- i\;e^{i\phi} \sin \frac{\theta}{2} \ket{0} + e^{2i\phi} \cos \frac{\theta}{2} \ket{1}. 
\end{aligned}
\end{equation}
From these two equations above we immediately see that the Hadamard operation is just the special case where $\theta = \phi = \pi/2$. 
\par
In QKD protocols based on entanglement swapping the basis transformation is usually applied onto one qubit of a Bell state. Taking the general transformation $T_x(\theta,\phi)$ from eq. (\ref{eq:OpTransXMatrix}) into account, the Bell state $\ket{\Phi^+}$ changes into
\begin{equation}
\begin{aligned}
T^{(1)}_x\bigl(\theta, \phi\bigr) \ket{\Phi^+}_{12} 
= &\cos \frac{\theta}{2} \; \frac{1}{\sqrt{2}} \Bigl( \ket{00} + e^{2i\phi} \ket{11} \Bigr) \\ 
- i\;e^{i\phi} &\sin \frac{\theta}{2} \; \frac{1}{\sqrt{2}} \Bigl( \ket{01} + \ket{10} \Bigr)
\label{eq:OpTransBellState}
\end{aligned}
\end{equation}
and accordingly for the other Bell states. The superscript ''$(1)$'' in eq. (\ref{eq:OpTransBellState}) indicates that the transformation $T_x\bigl(\theta, \phi\bigr)$ is applied on qubit 1. As a consequence, the application of $T_x(\theta,\phi)$ before the entanglement swapping is performed changes the results based on the angles $\theta$ and $\phi$. In detail, we have the state 
\begin{equation}
\begin{aligned}
T^{(1)}_x\bigl(\theta, \phi\bigr) \ket{\Phi^+}_{12} &\ket{\Phi^+}_{34} = \\
\frac{1}{2} \biggl(
  &\ket{\Phi^+}_{13} T^{(2)}_x\bigl(\theta, \phi\bigr) \ket{\Phi^+}_{24} \\ 
 +&\ket{\Phi^-}_{13} T^{(2)}_x\bigl(\theta, \phi\bigr) \ket{\Phi^-}_{24} \\[6pt]
 +&\ket{\Psi^+}_{13} T^{(2)}_x\bigl(\theta, \phi\bigr) \ket{\Psi^+}_{24} \\ 
 +&\ket{\Psi^-}_{13} T^{(2)}_x\bigl(\theta, \phi\bigr) \ket{\Psi^-}_{24}  
\biggr)
\end{aligned}
\label{eq:OpTransEntSwapping}
\end{equation}
upon which Alice performs her Bell state measurement on qubits 1 and 3 (cf. Figure \ref{fig:ES-QKD}). Here, the superscripts ''$(1)$'' and ''$(2)$'' in eq. (\ref{eq:OpTransEntSwapping}) indicate that after Alice's Bell state measurement on qubits 1 and 3 the transformation $T_x\bigl(\theta, \phi\bigr)$ swaps from qubit 1 onto qubit 2. When Bob performs his Bell state measurement on qubits 2 and 4, he obtains a result correlated to Alice's measurement outcome only with probability (cf. eq. (\ref{eq:OpTransBellState}) and eq. (\ref{eq:OpTransEntSwapping}) above)
\begin{equation}
P_{corr} = \cos^2\frac{\theta}{2} \; \cos^2(\phi).
\end{equation} 
Otherwise, he obtains an uncorrelated result, which differs from Alice's result by a Pauli operation, i.e. $\sigma_z$ or $\sigma_x$. That becomes a problem because Bob is no longer able to compute Alice's state based on his result and vice versa. 
\par
Fortunately, Bob can resolve this problem by transforming the state back into its original form. Following eq. (\ref{eq:OpTransEntSwapping}), where Alice performs $T_x\bigl(\theta, \phi\bigr)$ on qubit 1, he achieves that by applying the inverse $T^{-1}_x\bigl(\theta, \phi\bigr)$ on qubit 2 of his state. As we will see in the following section, if an adversary interferes with the communication, the effects of Alice's basis transformation can not be represented as in eq. (\ref{eq:OpTransEntSwapping}) any longer. Thus, even if Bob applies the inverse transformation, Alice's and Bob's results are uncorrelated to a certain amount. This amount is reflected in an error rate detected by Alice and Bob during post processing.

\section{Application of Simple Basis Transformations} \label{sec:SimpleBT}
In this section we want to give a short review on the results in \cite{SS12,SS13} dealing with the scenarios where Alice or Bob or both parties randomly apply a simplified version of the basis transformation described above. The simplification addresses the angle $\phi$, i.e. the rotation about the $Z$-axis. In the security discussions in \cite{SS12}, the angle $\phi$ is fixed at $\pi/2$ for reasons of simplicity. That means, the rotation about the $Z$-axis is constant at an angle of $\pi/2$ such that only the angle $\theta$ can be chosen freely. As already pointed out above, the Hadamard operation equals $T_x(\pi/2,\pi/2)$ (cf. eq. (\ref{eq:OpTransX})). Hence, the focus of this section is to analyze, whether the Hadamard operation is optimal when applied by one party and by both parties, respectively. To distinguish between the basis transformations applied by Alice and Bob, we discuss a setup where a basis transformation about an angle $\theta_A$ is applied by Alice and a transformation about an angle $\theta_B$ is applied by Bob, respectively (cf. Figure \ref{fig:ES-QKD}). 

\subsection{Application of a Single Transformation} \label{susec:SimpleSingTrans}
From the structure of $\ket{\delta}_{1QR4TU}$ given in eq. (\ref{eq:DeltaCorrelations}) we see that Eve is able to obtain full information about Alice's and Bob's secret whenever Alice and Bob do not apply any basis transformation, i.e. they use the state $\ket{\delta}$ Eve introduced into the protocol. For the first scenario, where only Alice randomly applies the basis transformation,  the overall state of the system after Eve's distribution of the state $\ket{\delta}_{P-U}$ can simply be described as 
\begin{equation}
\ket{\delta^\prime} = T^{(1)}_x(\theta_A,\tfrac{\pi}{2}) \ket{\delta}_{1QR4TU}
\end{equation}
where the superscript ''(1)'' again indicates that $T_x(\theta_A,\pi/2)$ is applied on qubit 1. When Eve sends qubits $R$ and $Q$ to Alice and Bob, respectively, the state after Alice's Bell state measurement on qubits 1 and $R$ is
\begin{equation}
\begin{aligned}
  \cos \frac{\theta_A}{2} \; \ket{\Phi^-}_{Q4} \ket{\varphi_2}_{TU} 
+ \sin \frac{\theta_A}{2} \; \ket{\Psi^+}_{Q4} \ket{\varphi_3}_{TU} 
\end{aligned}
\label{eq:DeltaPrimeTCorr}
\end{equation} 
assuming Alice obtained $\ket{\Phi^+}_{1R}$ (for Alice's other three possible results the state changes accordingly). Comparing this with eq. (\ref{eq:OpTransBellState}) and eq. (\ref{eq:OpTransEntSwapping}) it indicates that in this case Bob's transformation back into the $Z$-basis does not re-establish the correlations between Alice and Bob properly. Performing the calculations we see that Bob's operation $T_x(\theta_A,\pi/2)$ brings qubits $Q$, 4, $T$ and $U$ into the form
\begin{equation}
\begin{aligned}
  \ket{\Phi^+}_{Q4} &\biggl[ 
  \cos^2 \frac{\theta_A}{2} \; \ket{\varphi_2}_{TU} + \sin^2 \frac{\theta_A}{2} \; \ket{\varphi_3}_{TU}
  \biggr] \\
- \ket{\Psi^-}_{Q4} &\biggl[
  \frac{\sin\theta_A}{2} \; \ket{\varphi_2}_{TU} - \frac{\sin\theta_A}{2} \; \ket{\varphi_3}_{TU}
\biggr].
\end{aligned}
\label{eq:BobsState}
\end{equation}
When Bob performs a Bell state measurement, we can directly see from this expression that Bob obtains either the correlated result $\ket{\Phi^+}_{Q4}$ with probability
\begin{equation}
\begin{aligned}
P_{corr} &= 
   \biggl(\cos^2 \frac{\theta_A}{2}\biggr)^2 + \biggl(\sin^2 \frac{\theta_A}{2}\biggr)^2 \\
&= \frac{3+\cos(2\theta_A)}{4}
\end{aligned}
\label{eq:BobCorrResult} 
\end{equation}
or an error, i.e. the state $\ket{\Psi^-}_{Q4}$, otherwise. In detail, Eve introduces an error with probability $(\sin^2\theta_A)/2$ and since Alice applies the basis transformation at random, i.e. with probability 1/2, this yields an expected error probability
\begin{equation}
\langle P_e \rangle = \frac{1}{4} \sin^2\theta_A.
\label{eq:ExpErrorSingle}
\end{equation}
It is of major interest for the adversary Eve that Bob obtains a correlated result with a high probability such that her interference stays undetected. Nevertheless, based on the results in eq. (\ref{eq:BobsState}) we see that it is not possible for Eve to perform an operation on $T$ and $U$ such that Bob will obtain a correlated result with higher probability than given in eq. (\ref{eq:BobCorrResult}). This is due to the fact that Eve has to choose the states $\ket{\varphi_i}$ in a way that she is able to perfectly distinguish between them. Therefore, they have to be orthogonal, as already pointed out in eq. (\ref{eq:Varphi}) above.
\par
Besides staying undetected, Eve's second interest is to obtain as much information on the classical raw key bits as possible. The probability that Eve obtains the same classical bit from her measurement as Alice and Bob obtain from their Bell state measurements is called the \emph{collision probability} $P_c$. As already pointed out above, if Alice and Bob do not apply a basis transformation, Eve always has full information about Alice's and Bob's measurement outcomes, i.e. their classical raw key bits (cf. eq. (\ref{eq:DeltaCorrelations}) above). In the current scenario with one basis transformation $T_x(\theta_A,\pi/2)$, Eve obtains the state $\ket{\varphi_2}_{TU}$ from her Bell state measurement on qubits $T$ and $U$ with probability
\begin{equation}
P_c = \frac{\cos^2\frac{\theta}{2}}{\frac{3+\cos\;(2\theta_A)}{4}} = \frac{(1 + \cos(\theta_A))^2}{3 + \cos\;(2\theta_A)}
\end{equation}
as long as Alice's and Bob's results are correlated. This probability comes directly from eq. (\ref{eq:BobsState}) and eq. (\ref{eq:BobCorrResult}) above and in that case Eve knows that Bob obtained $\ket{\Phi^+}_{Q4}$ and further on the respective raw key bits. Summing over all possible measurement outcomes for Alice, Bob and Eve in all possible scenarios (i.e. application of $T_x(\theta_A,\pi/2)$ and no basis transformation at all) the expected collision probability for the whole protocol can be computed and results in
\begin{equation}
\langle P_c \rangle = \frac{1}{8} \Bigl( 7 + \cos 2\theta_A \Bigr).
\end{equation} 
This directly leads to the \emph{Shannon entropy}
\begin{equation}
H = \frac{1}{2} \; h\Bigl( \cos^2 \frac{\theta_A}{2} \Bigr)
\label{eq:ShannonEntropySingle}
\end{equation} 
where $h(x) = -x \log_2 x - (1-x) \log_2 (1-x)$ is the binary entropy. 
\par
\begin{figure}
\centering
\includegraphics[width=0.4\textwidth]{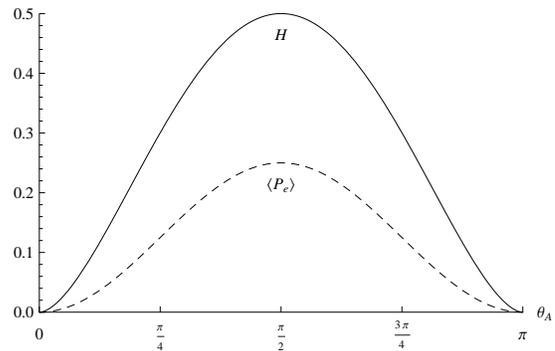}
\caption{Alice's and Bob's Shannon entropy $H$ and the according average error probability $\langle P_e \rangle$ if either Alice or Bob randomly applies a basis transformation.}
\label{fig:ShannonEntropy2D}
\end{figure}

 
Looking at the plot of $\langle P_e \rangle$ and $H$ in Figure \ref{fig:ShannonEntropy2D}, we see that the optimal angle $\theta_A$ for a single basis transformation is $\pi/2$, i.e. the Hadamard operation. Hence, for protocols using only one basis transformation, the application of the Hadamard operation is optimal, as it is already known from literature \cite{Cab01,SS08, SS12}. In this case the average error probability as well as the Shannon entropy are maximal at $\langle P_e \rangle = 0.25$ and $H = 0.5$ (cf. Figure \ref{fig:ShannonEntropy2D}). If only Bob applies the basis transformation the calculations run analogous and therefore provide the same results. Further, Eve's information on the bits of the secret key is given by the \emph{mutual information} 
\begin{equation}
I_{AE} = 1 - H = 1- \frac{1}{2} = \frac{1}{2}
\label{eq:MutualInfoSingle}
\end{equation}
which means that Eve has 0.5 bits of information on every bit of the secret key. Using post processing methods, i.e. error correction and privacy amplification, Eve's information can be brought below 1 bit of the whole secret key as long as the error rate is below $\sim 11\%$ \cite{SP00}. This is more or less the standard threshold value for the prepare and measure QKD protocols.

\subsection{Application of Combined Transformations} \label{susec:SimpleCombTrans}
In the previous paragraphs, we discussed the application of a basis transformation at either Alice's or Bob's side. When both parties apply a basis transformation with a different angle, i.e. $T_x(\theta_A,\pi/2)$ and $T_x(\theta_B,\pi/2)$, then the overall state changes to
\begin{equation}
\ket{\delta^\prime} = T^{(1)}_x(\theta_A,\tfrac{\pi}{2}) \; T^{(4)}_x(\theta_B,\tfrac{\pi}{2}) \; \ket{\delta}_{1QR4TU}.
\end{equation}
After Alice's Bell state measurement on qubits 1 and $R$ and Bob's application of $T_x(\theta_B,\pi/2)$ on qubit $Q$ the state of the remaining qubits is
\begin{equation}
\begin{aligned}
  \ket{\Phi^+}_{Q4} \biggl[
 &\cos^2 \frac{\theta_A-\theta_B}{2} \; \ket{\varphi_1}_{TU} \\
+&\sin^2 \frac{\theta_A-\theta_B}{2} \; \ket{\varphi_4}_{TU}
  \biggr] \\
- \ket{\Psi^-}_{Q4} \biggl[
 &\sin^2 \frac{\theta_A-\theta_B}{2} \; \ket{\varphi_1}_{TU} \\
-&\sin^2 \frac{\theta_A-\theta_B}{2} \; \ket{\varphi_4}_{TU}
  \biggr]
\end{aligned}
\label{eq:DeltaPrimeTCorrRev}
\end{equation}
Also in this case Eve measures either $\ket{\varphi_1}$ or $\ket{\varphi_4}$ and has no opportunity to perform an operation such that Bob obtains a result correlated to Alice's outcome. Consequently, Bob observes a correlated result with a probability similar to eq. (\ref{eq:BobCorrResult}) above, i.e.
\begin{equation}
\begin{aligned}
P_{corr} &= 
   \biggl(\cos^2 \frac{\theta_A - \theta_B}{2}\biggr)^2 + \biggl(\sin^2 \frac{\theta_A - \theta_B}{2}\biggr)^2 \\ 
&= \frac{3+\cos(2\theta_A - 2\theta_B)}{4}
\end{aligned}
\end{equation}
and obtains an error with probability $(\sin^2(\theta_A - \theta_B))/2$
\par
\begin{figure}
\centering
\includegraphics[width=0.4\textwidth]{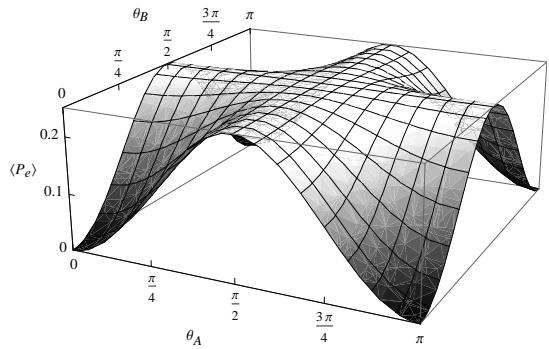}
\caption{The average error probability $\langle P_e \rangle$ if both parties randomly apply basis transformations described only by the angles $\theta_A$ and $\theta_B$.}
\label{fig:ErrProb3D}
\end{figure}

 
The average error probability for all possible scenarios, i.e. either no basis transformation, a single basis transformation by one party or the combination of two different basis transformations, is computed as the weighted sum over the error probabilities of these scenarios. Since Alice and Bob apply their basis transformations at random, each error probability is weighted with 1/4. This yields an average error probability for all scenarios (cf. Figure \ref{fig:ErrProb3D} for a plot of this function) 
\begin{equation}
\begin{aligned}
\langle P_e \rangle = \frac{1}{8} &\sin^2 \theta_A 
					+ \frac{1}{8}  \sin^2 \theta_B \\
					+ \frac{1}{16} &\sin^2 \bigl(\theta_A + \theta_B\bigr) 
					+ \frac{1}{16}  \sin^2 \bigl(\theta_A - \theta_B\bigr).
\end{aligned}
\label{eq:ExpErrorCombined} 
\end{equation}
\par
\begin{figure}
\centering
\includegraphics[width=0.4\textwidth]{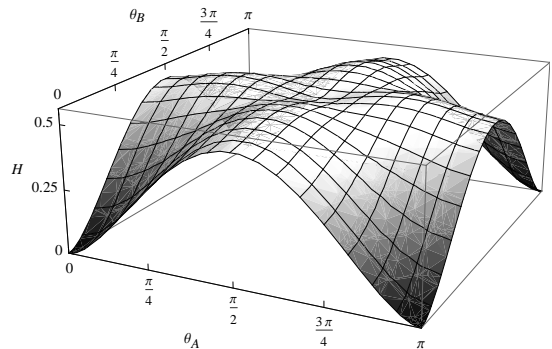}
\caption{Alice's and Bob's Shannon entropy $H$ if both parties randomly apply 
basis transformations described only by the angles $\theta_A$ and $\theta_B$.}
\label{fig:ShannonEntropy3D}
\end{figure}

When the results are correlated, Eve obtains either $\ket{\varphi_1}_{TU}$ or $\ket{\varphi_4}_{TU}$, as it can be computed from eq. (\ref{eq:DeltaPrimeTCorrRev}). Based on this fact, the collision probability and further on Alice's and Bob's Shannon entropy can be computed following the same argumentation as described in the other scenarios. The Shannon entropy of all for scenarios is again computed as the weighted sum of the single Shannon entropies (cf. eq. (\ref{eq:ShannonEntropySingle})). Due to the interference term resulting from the application of two different basis transformations, the Shannon entropy is higher compared to a single basis transformation (cf. Figure \ref{fig:ShannonEntropy3D} for a plot of this function):
\begin{equation}
\begin{aligned}
H = &\frac{1}{4} \; h\Bigl( \cos^2 \frac{\theta_A}{2} \Bigr) 
   + \frac{1}{4} \; h\Bigl( \cos^2 \frac{\theta_B}{2} \Bigr) \\
  + &\frac{1}{8} \; h\Bigl( \cos^2 \frac{\theta_A + \theta_B}{2} \Bigr) 
  +  \frac{1}{8} \; h\Bigl( \cos^2 \frac{\theta_A - \theta_B}{2} \Bigr). 
\end{aligned}
\label{eq:ShannonEntropyCombined}
\end{equation}
This is due to the fact that it is more difficult for Eve to react on two separate basis transformations with different angles $\theta_A$ and $\theta_B$. Further, we see that eq. (\ref{eq:ShannonEntropyCombined}) reduces to eq. (\ref{eq:ShannonEntropySingle}) if either $\theta_A = 0$ or $\theta_B = 0$. Hence, taking the optimal choice for only one basis transformation, i.e. the Hadamard operation, we see that if both parties apply the Hadamard operation at the same time, the operations cancel out each other. Therefore, a requirement to reach a better result is that the angles $\theta_A$ and $\theta_B$ have to be different. As we can further see from Figure \ref{fig:ShannonEntropy3D}, the Shannon entropy for a combined application of basis transformations is higher than 0.5 for some regions. In detail, the maximum of the function plotted in Figure \ref{fig:ShannonEntropy3D} is 
\begin{equation}
H \sim 0.55 \quad \text{and thus} \quad I_{AE} \sim 0.45
\label{eq:MutualInfoCombined}
\end{equation}
for $\theta_A \in \{\pi/4,3\pi/4\}$  and $\theta_B = \pi/2$ or vice versa. Hence, if just one of the parties applies a Hadamard operation and the other one a transformation about an angle of $\pi/4$ or $3\pi/4$, Eve's mutual information is about 10\% lower compared to the application of a single basis transformation (cf. eq. (\ref{eq:MutualInfoSingle})). At the same time, we see from Figure \ref{fig:ErrProb3D} that for these two values of $\theta_A$ and $\theta_B$ the error probability is still maximal with $\langle P_e \rangle = 0.25$. This means, Alice and Bob are able to further increase the security by the combined application of two basis transformations.

\section{Application of General Basis Transformations} \label{sec:GeneralBT}
In the following paragraphs, we want to extend the results from the previous section by applying general basis transformations, which means Alice and Bob are able to choose both angles $\theta$ and $\phi$ in eq. (\ref{eq:OpTransX}) freely. Following the scenarios from the previous section, we are at first looking only on one party performing a basis transformation on their respective qubits and afterwards on two different basis transformations performed by each of the parties. For each scenario we will show, which values for $\theta$ and $\phi$ are optimal to give an adversary the least information about the raw key bits. In the course of the two scenarios, we will denote Alice's operation as $T_x\bigl(\theta_A, \phi_A\bigr)$ and accordingly Bob's operation as $T_x\bigl(\theta_B, \phi_B\bigr)$.

\subsection{Application of a Single Transformation} \label{susec:GeneralSingleBT}
As already pointed out above, the application of the basis transformation occurs at random and Eve is able to obtain full information about Alice's and Bob's secret due to the structure of the state $\ket{\delta}$, if the two parties do not apply any basis transformation at all. Therefore, we look at first at the effects of a basis transformation at Alice's side. Her initial application of the general basis transformation $T_x\bigl(\theta_A, \phi_A\bigr)$ does alter the state $\ket{\delta}_{1QR4TU}$ introduced by Eve such that it is changed to
\begin{equation}
\ket{\delta^\prime}_{1QR4TU} = T^{(1)}_x\bigl(\theta_A, \phi_A\bigr) \ket{\delta}_{1QR4TU}
\label{eq:DeltaPrime1}
\end{equation} 
After a little algebra, we see that Alice obtains all four Bell states with equal probability and after her measurement the state of the remaining qubits is 

\begin{equation}
\begin{aligned}
    e^{i \phi_A} \cos \frac{\theta_A}{2} \cos \phi_A \; &\ket{\Phi^+}_{Q4} \ket{\varphi_1}_{TU} \\
- i e^{i \phi_A} \cos \frac{\theta_A}{2} \sin \phi_A \; &\ket{\Phi^-}_{Q4} \ket{\varphi_2}_{TU} \\
- i e^{i \phi_A} \sin \frac{\theta_A}{2}           \; &\ket{\Psi^+}_{Q4} \ket{\varphi_3}_{TU} 
\end{aligned}
\label{eq:DeltaPrime1Corr}
\end{equation}
assuming Alice obtained $\ket{\Phi^+}_{1R}$. We are presenting just the state for this particular result in detail because it would be simply too complex to present the representation of the whole state for all possible outcomes here. Nevertheless, for the other three possible results the remaining qubits end up in a similar state, where only Bob's Bell states of the qubits $Q$ and 4 as well as Eve's auxiliary states of the qubits $T$ and $U$ change accordingly to Alice's measurement result. 
\par 
Before Bob performs his Bell state measurement, he first has to reverse Alice's basis transformation. This can be achieved by applying $T^{-1}_x\bigl(\theta_A, \phi_A\bigr)$ on qubit $Q$ in his possession. Whereas this would reverse the effect of Alice's basis transformation if no adversary is present, the structure of Eve's state $\ket{\delta}$ makes this reversion impossible, as already pointed out in the previous section. Therefore, the state in eq. (\ref{eq:DeltaPrime1Corr}) changes into
\begin{widetext}
\begin{equation}
\begin{aligned}
  e^{i \phi_A} \cos^2 \frac{\theta_A}{2} \biggl[
    \cos \phi_A \; &\frac{1}{\sqrt{2}} \Bigl( \ket{00}_{Q4} + e^{-2i \phi_A} \ket{11}_{Q4} \Bigr) \ket{\varphi_1}_{TU} 
 -i \sin \phi_A \; \frac{1}{\sqrt{2}} \Bigl( \ket{00}_{Q4} - e^{-2i \phi_A} \ket{11}_{Q4} \Bigr) \ket{\varphi_2}_{TU} \\
 +e^{-i \phi_A} \tan^2 \frac{\theta_A}{2} \; &\frac{1}{\sqrt{2}} \Bigl( \ket{00}_{Q4} + \ket{11}_{Q4} \Bigr) \ket{\varphi_3}_{TU} 
  \biggr] \\
+ i \cos \frac{\theta_A}{2} \sin \frac{\theta_A}{2} \biggl[
    \cos \phi_A \; &\frac{1}{\sqrt{2}} \Bigl( \ket{00}_{Q4} + \ket{11}_{Q4} \Bigr) \ket{\varphi_1}_{TU} 
+ i \sin \phi_A \; \frac{1}{\sqrt{2}} \Bigl( \ket{00}_{Q4} - \ket{11}_{Q4} \Bigr) \ket{\varphi_2}_{TU} \\
- e^{i \phi_A} &\frac{1}{\sqrt{2}} \Bigl( \ket{01}_{Q4} + e^{-2i \phi_A} \ket{10}_{Q4} \Bigr) \ket{\varphi_3}_{TU}
\biggr]
\end{aligned}
\label{eq:DeltaPrime1CorrRev}
\end{equation}
\end{widetext}
for Alice's result $\ket{\Phi^+}_{1R}$ and accordingly for the other results. Therefore, Bob obtains the correlated state $\ket{\Phi^+}_{Q4}$ only with probability
\begin{equation}
P_{\Phi^+} = \frac{1}{4} \Bigl( 3 + \cos \bigl(4 \phi_A \bigr) \Bigr) \cos^4 \frac{\theta_A}{2} + \sin^4 \frac{\theta_A}{2}
\label{eq:ProbCorrBob1}
\end{equation}
and the other results with probability
\begin{equation}
\begin{aligned}
P_{\Phi^-} &= 2 \cos^4 \frac{\theta_A}{2} \cos^2 \phi_A \sin^2 \phi_A \\
P_{\Psi^+} &= \frac{1}{2} \sin^2 \theta_A \cos^2 \phi_A \\
P_{\Psi^-} &= \frac{1}{2} \sin^2 \theta_A \sin^2 \phi_A.
\end{aligned}
\end{equation} 
Hence, due to Eve's intervention Bob obtains a result uncorrelated to Alice's outcome with probability
\begin{equation}
\begin{aligned}
P_e &= P_{\Phi^-} + P_{\Psi^+} + P_{\Psi^-} \\
    &= \frac{1}{2} \Bigl(\sin^2 \theta_A + \cos^4 \frac{\theta_A}{2} \sin^2 \bigl(2\phi_A\bigr) \Bigr).
\end{aligned}
\label{eq:ProbErrBob1}
\end{equation}
Assuming that Bob obtains $\ket{\Phi^+}_{Q4}$, i.e. the expected result based on Alice's measurement outcome, Eve obtains either $\ket{\varphi_1}$, $\ket{\varphi_2}$ or $\ket{\varphi_3}$ from her measurement on qubits $T$ and $U$ in the state described in eq. (\ref{eq:DeltaPrime1CorrRev}) with the respective probabilities 
\begin{equation}
\begin{aligned}
P_{\varphi_1} &= \frac{\cos^4 \frac{\theta_A}{2} \cos^4 \phi_A}{\frac{1}{4}(3 + \cos 4 \phi_A) \cos^4 \frac{\theta_A}{2} + \sin^4 \frac{\theta_A}{2}} \\
P_{\varphi_2} &= \frac{\cos^4 \frac{\theta_A}{2} \sin^4 \phi_A}{\frac{1}{4}(3 + \cos 4 \phi_A) \cos^4 \frac{\theta_A}{2} + \sin^4 \frac{\theta_A}{2}} \\
P_{\varphi_3} &= \frac{-\sin^2 \frac{\theta_A}{2}}{(3 + \cos 4 \phi_A) \cos^4 \frac{\theta_A}{2} + 4 \sin^4 \frac{\theta_A}{2}}
\end{aligned}
\label{eq:ProbEve1}
\end{equation}
Furthermore, in case Bob measures an uncorrelated result, Eve obtains two out of the four auxiliary states $\ket{\varphi_i}$ at random. Hence, due to the basis transformation $T_x\bigl(\theta_A, \phi_A\bigr)$, Eve's auxiliary systems are less correlated to Bob's result compared to the application of a simple basis transformation as described in eq. (\ref{eq:BobsState}) and eq. (\ref{eq:DeltaPrimeTCorrRev}) above. In other words, Eve's information on Alice's and Bob's result is further reduced compared to the previous scenarios.
\par
Since Alice applies the basis transformation at random, i.e. with probability 1/2, the average error probability $\langle P_e \rangle$ can be directly computed using eq. (\ref{eq:ProbErrBob1}) and its variations based on Alice's measurement result as
\begin{equation}
\begin{aligned}
\langle P_e \rangle = \frac{1}{4} \biggl[
 & \sin^2 \theta_A + \cos^4 \frac{\theta_A}{2} \sin^2 \Bigl(2 \phi_A\Bigr)
\biggr].
\end{aligned} 
\label{eq:ExpErrProbGeneral1}
\end{equation}
\begin{figure}
\centering
\includegraphics[width=0.4\textwidth]{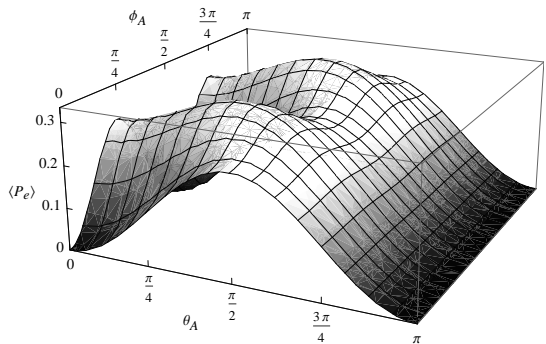}
\caption{The average error probability $\langle P_e \rangle$ depending on the angles $\theta_A$ and $\phi_A$}
\label{fig:ExpErrProbGeneral1}
\end{figure}
Keeping in mind that Eve does not introduce any error when Alice does not use the basis transformation $T_x\bigl(\theta_A, \phi_A\bigr)$, the average collision probability $\langle P_c \rangle$ can be computed as (cf. also eq. (\ref{eq:ProbEve1}))
\begin{equation}
\begin{aligned}
\langle P_c \rangle = \frac{1}{64} \Bigl(
53 &- 4\cos \theta_A + 7\cos \bigl(2\theta_A\bigr) \\
   &+ 8 \cos^4 \frac{\theta_A}{2}\cos \bigl(4 \phi_A\bigr) 
\Bigr).
\end{aligned}
\label{eq:ExpCollProbGeneral1}
\end{equation}
In further consequence this leads to the Shannon entropy $H$ of the raw key, i.e.  
\begin{equation}
H = \frac{1}{2} \biggl[ 
h \Bigl( \cos^2 \frac{\theta_A}{2} \Bigr) + \cos^2 \frac{\theta_A}{2}\; h \Bigl( \cos^2 \phi_A \Bigr)
\biggr].
\label{eq:ShannonEntropyGeneral1}
\end{equation}
\par
\begin{figure}
\centering
\includegraphics[width=0.4\textwidth]{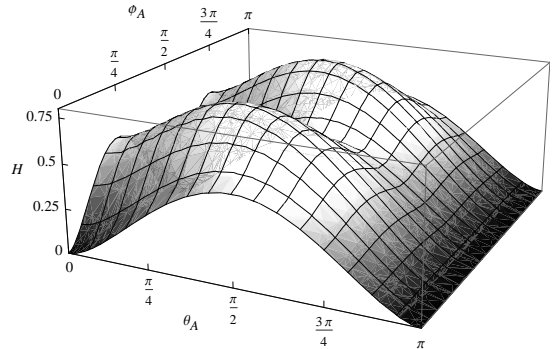}
\caption{Alice's and Bob's Shannon entropy $H$ of the raw key depending on the angles $\theta_A$ and $\phi_A$}
\label{fig:ShannonEntropyGeneral1}
\end{figure}

As we can directly see from Figure \ref{fig:ExpErrProbGeneral1}, the average error probability $\langle P_e \rangle$ has its maximum at 1/3 with $\theta_A \simeq 0.39183 \pi$ and $\phi_A = \pi/4$ or $\phi_A = 3\pi/4$. For this choice of $\theta_A$ and $\phi_A$ we see from Figure \ref{fig:ShannonEntropyGeneral1} that the Shannon entropy is also maximal with $H \simeq 0.79248$. Hence, the adversary Eve is left with a mutual information of
\begin{equation}
I_{AE} = 1 - H = 0.20752
\end{equation}
This value for the mutual information is less than half of Eve's information on the raw key compared to the application of a simplified basis transformation (cf. eq. (\ref{eq:MutualInfoSingle}) and eq. (\ref{eq:MutualInfoCombined}) above). Hence, the application of a general basis transformation by either Alice or Bob decreases Eve's information on the raw key and simultaneously increases the probability to detect the eavesdropping attempt compared to the application of one or two simple basis transformations. 
\par
Unfortunately, the angle for $\theta_A \simeq 0.39183\pi$ to reach the maximum value is rather odd and difficult to realize in a practical implementation. In contrast, an angle $\theta_A = 3 \pi/8$ is more convenient and much easier to realize. For this scenario we can compute from eq. (\ref{eq:ExpErrProbGeneral1}) an average error rate of $\langle P_e \rangle \simeq 0.33288$ and from eq. (\ref{eq:ShannonEntropyGeneral1}) the respective Shannon entropy $H \simeq 0.79148$ (cf. also  Figure \ref{fig:ExpErrProbGeneral1} and Figure \ref{fig:ShannonEntropyGeneral1}), which are both just insignificantly lower than their maximum values. Accordingly, Eve's mutual information on the raw key is slightly above 20\%, i.e. $I_{AE} \simeq 0.20852$. 
Hence, the security of the protocol is drastically increased using a general basis transformation compared to the application of a Hadamard operation.  
 
\subsection{Application of Combined Transformations} \label{susec:GeneralDoubleBT}
In the previous paragraphs, we discussed the application of one general basis transformation $T_x\bigl(\theta_A, \phi_A\bigr)$ on Alice's side. It is easy to see that the results for the average error probability $\langle P_e \rangle$ in eq. (\ref{eq:ExpErrProbGeneral1}) as well as the Shannon entropy $H$ in eq. (\ref{eq:ShannonEntropyGeneral1}) are the same if only Bob randomly applies the basis transformation $T_x\bigl(\theta_B, \phi_B\bigr)$ on his side. 
\par 
Hence, the more interesting scenario is the combined random application of two different basis transformations $T_x\bigl(\theta_A, \phi_A\bigr)$ on Alice's side and $T_x\bigl(\theta_B, \phi_B\bigr)$ on Bob's side. The application of these two different basis transformations alters the initial state accordingly to
\begin{equation}
\ket{\delta^\prime}_{1QR4TU} = T^{(1)}_x\bigl(\theta_A, \phi_A\bigr) \; T^{(4)}_x\bigl(\theta_B, \phi_B\bigr) \; \ket{\delta}_{1QR4TU}
\label{eq:DeltaPrime2}
\end{equation} 
where again the superscripts ''(1)'' and ''(4)'' indicate that $T_x\bigl(\theta_A, \phi_A\bigr)$ is applied on qubit 1 and $T_x\bigl(\theta_B, \phi_B\bigr)$ on qubit 4, respectively. Following the protocol, Alice has to undo Bob's transformation using $T^{-1}_x\bigl(\theta_B, \phi_B\bigr)$ before she can perform her Bell state measurement. Similar to the application of one basis transformation described above, Alice obtains all four Bell states with equal probability from her measurement. The state of the remaining qubits changes in a way analogous to eq. (\ref{eq:DeltaPrime1Corr}) above and Bob has to reverse Alice's transformation using $T^{-1}_x\bigl(\theta_A, \phi_A\bigr)$. This changes the state similar to eq. (\ref{eq:DeltaPrime1CorrRev}) above. Hence, when Bob performs his measurement on qubits $Q$ and 4, he does not only obtain a result correlated to Alice's outcome, but all four possible Bell states with different probabilities such that an error is introduced in the protocol. As already discussed in the previous section, the results from Eve's measurement on qubits $T$ and $U$ are not fully correlated to Alice's and Bob's results and therefore Eve's information on the raw key bits is further reduced compared to the application of only one transformation.    
\par 
Due to the fact that Alice as well as Bob choose at random whether they apply their respective basis transformation, the average error probability is calculated over all scenarios, i.e. no transformation is applied, either Alice or Bob applies $T_x\bigl(\theta_A, \phi_A\bigr)$ and $T_x\bigl(\theta_B, \phi_B\bigr)$, respectively, or both transformations are applied. Therefore, using the results from eq. (\ref{eq:ExpErrProbGeneral1}) above, the overall error probability can be computed as
\begin{equation}
\begin{aligned}
\langle P_e \rangle = 
  \frac{1}{8} \biggl[ &\sin^2 \theta_A + \cos^4 \frac{\theta_A}{2} \sin^2 \Bigl(2 \phi_A\Bigr) \biggr] \\
+ \frac{1}{8} \biggl[ &\sin^2 \theta_B + \cos^4 \frac{\theta_B}{2} \sin^2 \Bigl(2 \phi_B\Bigr) \biggr] \\
+ \frac{1}{16} \biggl[ &\sin^2 \Bigl( \theta_A + \theta_B \Bigr) \\
+ &\cos^4 \frac{\theta_A + \theta_B}{2} \sin^2 \Bigl(2 \bigl(\phi_A + \phi_B\bigr)\Bigr) \biggr] \\
+ \frac{1}{16} \biggl[ &\sin^2 \Bigl( \theta_A - \theta_B \Bigr) \\
+ &\cos^4 \frac{\theta_A - \theta_B}{2} \sin^2 \Bigl(2 \bigl(\phi_A - \phi_B\bigr)\Bigr) \biggr]
\end{aligned}
\label{eq:ExpErrProbGeneral2}
\end{equation}
\begin{figure}
\centering
\includegraphics[width=0.4\textwidth]{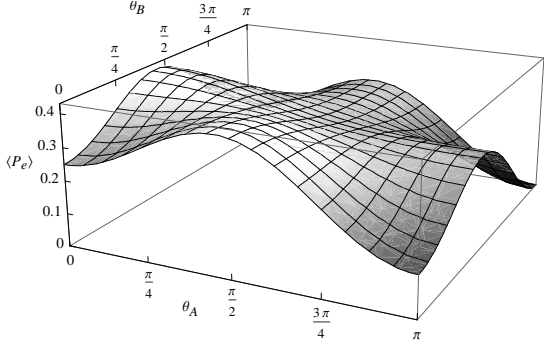}
\caption{Eve's expected error probability $\langle P_e \rangle$ depending on the angles $\theta_A$ and $\theta_B$. The remaining parameters $\phi_A$ and $\phi_B$ are fixed at $\pi/4$.}
\label{fig:ExpErrProbGeneral2}
\end{figure}
having its maximum at $\langle P_e \rangle \simeq 0.41071$. One possibility to reach the maximum is to choose the angles
\begin{equation}
\begin{aligned}
\theta_A = 0 &\qquad \theta_B \simeq 0.45437\pi \\ 
\phi_A = \frac{\pi}{4} &\qquad \phi_B = \frac{\pi}{4}.
\end{aligned}
\end{equation}
In fact, as long as $\phi_A =\pi/4$ or $\phi_A =3\pi/4$ the value of $\phi_B$ can be chosen freely to reach the maximum. Hence, the average error probability is plotted in Figure \ref{fig:ExpErrProbGeneral2} taking $\phi_A = \phi_B = \pi/4$.
\par
Following the same argumentation and using eq. (\ref{eq:ShannonEntropyGeneral1}) from above, the Shannon entropy can be calculated as
\begin{equation}
\begin{aligned}
H = 
  \frac{1}{4} \biggl[ &h\Bigl(\cos^2 \frac{\theta_A}{2} \Bigr) + \cos^2 \frac{\theta_A}{2} \; h\Bigl( \cos^2 \phi_A \Bigr) \biggr] \\
+ \frac{1}{4} \biggl[ &h\Bigl(\cos^2 \frac{\theta_B}{2} \Bigr) + \cos^2 \frac{\theta_B}{2} \; h\Bigl( \cos^2 \phi_B \Bigr) \biggr] \\
+ \frac{1}{8} \biggl[ &h\Bigl(\cos^2 \frac{\theta_A + \theta_B}{2} \Bigr) \\ 
+ &\cos^2 \frac{\theta_A + \theta_B}{2} \; h\Bigl( \cos^2 \bigl( \phi_A + \phi_B \bigr) \Bigr) \biggr] \\
+ \frac{1}{8} \biggl[ &h\Bigl(\cos^2 \frac{\theta_A - \theta_B}{2} \Bigr) \\ 
+ &\cos^2 \frac{\theta_A - \theta_B}{2} \; h\Bigl( \cos^2 \bigl( \phi_A - \phi_B \bigr) \Bigr) \biggr]
\end{aligned}
\label{eq:ShannonEntropyGeneral2}
\end{equation}
having its maximum at $H \simeq 0.9452$. This maximum is reached, for example, using 
\begin{equation}
\begin{aligned}
\theta_A \simeq -0.18865 \pi &\qquad \theta_B \simeq 0.42765 \pi \\
\phi_A   \simeq -0.22405 \pi &\qquad \phi_B \simeq 0.36218 \pi.
\end{aligned}
\end{equation}
The maximal Shannon entropy can also be reached using other values but they are not as nicely distributed as in the case of the average error probability.
\begin{figure}
\centering
\includegraphics[width=0.4\textwidth]{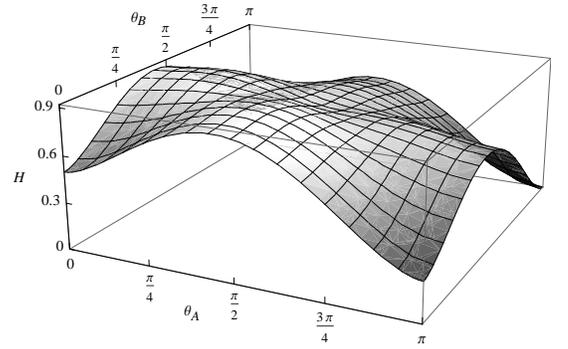}
\caption{Alice's and Bob's Shannon entropy $H$ of the raw key depending on the angles $\theta_A$ and $\theta_B$.  The remaining parameters $\phi_A$ and $\phi_B$ are fixed at $\pi/4$.}
\label{fig:ShannonEntropyGeneral2}
\end{figure}
\par
Since the maximal Shannon entropy is reached with a different set of values for $\theta_{\{A,B\}}$ and $\phi_{\{A,B\}}$ than the maximal error probability, it is necessary to find a set of parameters for the basis transformation, which results in a high error probability as well as a high Shannon entropy at the same time. On the one hand, inserting the values to reach the maximal Shannon entropy in eq. (\ref{eq:ExpErrProbGeneral2}) we obtain an average error probability $\langle P_e \rangle \simeq 0.3993$. On the other hand, using the parameters for a maximal error probability in eq. (\ref{eq:ShannonEntropyGeneral2}) we get $H \simeq 0.90635$. Therefore, Alice and Bob have to decide, whether they want to maximize $\langle P_e \rangle$ or $H$, being aware that the respective other value  is not optimal any more.
\par
Looking again at set of values for $\theta_{\{A,B\}}$ and $\phi_{\{A,B\}}$, which are more suitable for a physical implementation than the values mentioned above, one possibility for Alice and Bob is to choose
\begin{equation}
\begin{aligned}
\theta_A = -\frac{3\pi}{16} &\qquad \theta_B = \frac{7\pi}{16} \\
\phi_A = -\frac{\pi}{4} &\qquad \phi_B = \frac{3\pi}{8}
\end{aligned}
\label{eq:Parameters1}
\end{equation} 
leading to an almost optimal Shannon entropy $H \simeq 0.9399$ and a average respective error probability $\langle P_e \rangle \simeq 0.39288$. Keeping $\phi_A$ and $\phi_B$ fixed -- as already discussed in the previous section -- such that
\begin{equation}
\begin{aligned}
\theta_A = \frac{3\pi}{16} &\qquad \theta_B = \frac{7\pi}{16} \\
\phi_A = \frac{\pi}{4} &\qquad \phi_B = \frac{\pi}{4}
\end{aligned}
\label{eq:Parameters2}
\end{equation}  
the same average error probability $\langle P_e \rangle \simeq 0.39288$ and a slightly smaller Shannon entropy $H \simeq 0.91223$ compared to the previous values are achieved. Hence, we see that using a set of parameters more suitable for a physical implementation still results in a high error rate and leaves Eve's mutual information $I_{AE}$ below 10\%. 

\section{Results and Implications}

The results presented in the previous sections have direct implications on the security of QKD protocols based on entanglement swapping. Where in some QKD protocols \cite{Cab01,Son04,LWWSZ06} a random application of a Hadamard operation is used to detect an eavesdropper and secure the protocol, the above results indicate that the Hadamard operation is not the optimal choice. Using the Hadamard operation leaves an adversary with a mutual information $I_{AE} = 0.5$ and an expected error probability $\langle P_e \rangle = 0.25$ (cf. Table \ref{tab:Overview}), which is comparable to standard prepare and measure protocols \cite{BB84,Eke91,BBM92}. An application of the Hadamard operation by both communication parties does not increase these values since the interference terms in eq. (\ref{eq:ExpErrorCombined}) and eq. (\ref{eq:ShannonEntropyCombined}) cancel out each other (cf. also Figure \ref{fig:ErrProb3D}).
\par 
A slight change in one of the basis transformations, e.g. from $\theta_B = \pi/2$ (in case of the Hadamard operation) to $\theta_B = \pi/4$ is enough to decrease the adversary's information by about 10\% to $I_{AE} \simeq 0.45$ while leaving the expected error probability at the same value of $\langle P_e \rangle = 0.25$ (cf. eq. (\ref{eq:ShannonEntropyCombined}) and Table \ref{tab:Overview}). This gives Alice and Bob a slight advantage compared to other entanglement swapping based protocols like \cite{Cab01,Son04,LWWSZ06}. In particular, looking at the protocol by Song \cite{Son04}, where a basis transformation about an angle $\theta_A = 2\pi/3$ is applied, Eve's information is reduced by almost 25\% from $I_{AE} \simeq 0.594$  (coming directly from eq. (\ref{eq:ShannonEntropySingle}) above) in the original protocol to $I_{AE} \simeq 0.45$ (cf. also \cite{SS13} for details). 
\par 
Giving Alice an additional degree of freedom, i.e. choosing both $\theta_A$ and $\phi_A$ freely, she is able to further decrease the adversary's information about the raw key bits. By shifting $\phi_A$ from $\pi/2$ to $\pi/4$ and $\theta_A$ from $\pi/2$ or $\pi/4$ to $3\pi/8$ the adversary's information is reduced to $I_{AE} \simeq 0.208$ (cf. eq. (\ref{eq:ShannonEntropyGeneral1})). This is a reduction by almost 60\% compared to QKD schemes described in \cite{BB84,Eke91,BBM92,Cab01,LWWSZ06} and more than 50\% compared to the combined application of two different basis transformations (cf. also \cite{SS12,SS13}). At the same time, the expected error probability is increased by one third to $\langle P_e \rangle \simeq 0.333$ (cf. eq. (\ref{eq:ExpErrProbGeneral1})). Hence, an adversary does not only obtain fewer information about the raw key bits but also introduces more errors and therefore is easier detectable.   
\begin{table}
\scriptsize
\begin{tabular}{|c||c|c|c|}
\hline  
& $\phi_A = 0$ & $\phi_A = \frac{\pi}{2}$ & $\phi_A = \frac{\pi}{4}$ \\[0.1cm] \hline \hline  
$\phi_B = 0$             & 
\begin{tabular}{c}
$\theta_A = 0$, $\theta_B = 0$ \\[0.1cm] \hline 
$\langle P_e \rangle = 0$ \\ 
$I_{AE} = 1$ \\ 
\end{tabular} & 
\begin{tabular}{c}
$\theta_A = \frac{\pi}{2}$, $\theta_B = 0$ \\[0.1cm] \hline
$\langle P_e \rangle = 0.25$ \\ 
$I_{AE} = 0.5 $ \\ 
\end{tabular} & 
\begin{tabular}{c}
$\theta_A = \frac{3 \pi}{8}$, $\theta_B = 0$ \\[0.1cm] \hline
$\langle P_e \rangle \simeq 0.333$ \\ 
$I_{AE} \simeq  0.208$ \\ 
\end{tabular} \\ 
\hline  
$\phi_B = \frac{\pi}{2}$ &  & 
\begin{tabular}{c}
$\theta_A = \frac{\pi}{2}$, $\theta_B = \frac{\pi}{4}$ \\[0.1cm] \hline
$\langle P_e \rangle = 0.25$\\ 
$I_{AE} \simeq 0.45 $ \\ 
\end{tabular} & 
\begin{tabular}{c}
$\theta_A = 0$, $\theta_B = \frac{\pi}{2}$ \\[0.1cm] \hline
$\langle P_e \rangle \simeq 0.334$ \\ 
$I_{AE} = 0.125$ \\ 
\end{tabular} \\ 
\hline  
$\phi_A = \frac{\pi}{4}$ &  &  & 
\begin{tabular}{c}
$\theta_A = \frac{3\pi}{16}$, $\theta_B = \frac{7\pi}{16}$ \\[0.1cm] \hline
$\langle P_e \rangle = 0.393$\\ 
$I_{AE} = 0.088 $ \\ 
\end{tabular} \\ 
\hline 
\end{tabular} 
\caption{Overview of the expected error rate $\langle P_e \rangle$ and Eve's information $I_{AE}$ on the raw key bits for different values of the angles $\theta_{A,B}$ and $\phi_{A,B}$.}
\label{tab:Overview}
\end{table}
\par 
Following these arguments, the best strategy for Alice and Bob is to apply different basis transformations at random to reduce the adversary's information to a minimum. As already pointed out above, this minimum of $I_{AE} \simeq 0.0548$ is reached with a rather odd configuration for $\theta_{\{A,B\}}$ and $\phi_{\{A,B\}}$ as described in the previous section. Hence, it is important to look at configurations more suitable for physical implementations, i.e. configurations of $\theta_{\{A,B\}}$ and $\phi_{\{A,B\}}$ described by simpler fractions of $\pi$ as given in eq. (\ref{eq:Parameters1}) and (\ref{eq:Parameters2}). In this case, we showed that $\phi_{\{A,B\}}$ can be fixed at $\phi_A = \phi_B = \pi/4$ and with $\theta_A = 3\pi/16$ and $\theta_B = 7\pi/16$ almost maximal values can be achieved resulting in $I_{AE} \simeq 0.088$ and $\langle P_e \rangle \simeq 0.393$ (cf. eq. (\ref{eq:Parameters2}) and also Table \ref{tab:Overview}). 
\par 
Regarding physical implementations, another -- even simpler -- configuration can be found, involving only $\pi/2$ and $\pi/4$ rotations (cf. Table \ref{tab:Overview}). In this case $\theta_A = 0$, $\phi_A = \pi/4$ and $\theta_B = \phi_B = \pi/2$ which leaves the expected error probability at $\langle P_e \rangle \simeq 0.334$. The adversary's information is nowhere near the minimum but still rather low at $I_{AE} = 0.125$.   
\par 
In terms of security these results represent a huge advantage over QKD protocols based on entanglement swapping \cite{Cab01,Son04,LWWSZ06} or standard prepare and measure protocols \cite{BB84,Eke91,BBM92}. Such protocols usually have an expected error probability of $\langle P_e \rangle = 0.25$ and a mutual information $I_{AE} = 0.5$. Due to the four degrees of freedom, the error rate is between one third ($\langle P_e \rangle \simeq 0.334$) and more than one half ($\langle P_e \rangle = 0.411$) higher in the scenarios described here than in the standard protocols, which makes it easier to detect an adversary. Taking the practical threshold value of about 11\% for the error rate \cite{SP00} into account, it is easy to see that an adversary has to attack fewer qubits in transit to stay below this threshold compared to standard prepare and measure protocols. Further, the adversary's information is reduced by 75\% (to $I_{AE} = 0.125$) in a conservative setting, and up to almost 90\% (to $I_{AE} = 0.055$) in the optimal setting. This makes it possible to allow a higher threshold value than 11\% for entanglement swapping based QKD protocols.

\section{Conclusion}

In this article, we discussed the effects of basis transformations on the security of quantum key distribution protocols based on entanglement swapping. We showed that the Hadamard operation, a transformation from the $Z$- into the $X$-basis often used in prepare and measure protocols, is not optimal in connection with entanglement swapping based protocols. Starting from a general basis transformation described by two angles $\theta$ and $\phi$, we inspected the effects on the security when the adversary follows a collective attack strategy. We showed that the application of a basis transformation by one of the communication parties decreases the adversary's information to about $I_{AE} \simeq 0.2075$, which is less than half of the information compared to an application of the Hadamard operation. At the same time, the average error probability introduced by the presence of the adversary increases to $\langle P_e \rangle = 1/3$. A combined application of two different basis transformations further reduces the adversary's information to about $I_{AE} \simeq 0.0548$ at an average error probability of slightly more than 0.41. 
\par
Since the configuration of the angles $\theta$ and $\phi$ to reach these maximal values is not very suitable for a physical implementation, we also showed that these maximal values are almost reached with more convenient values for $\theta$ and $\phi$. In this case, the adversary's information is still $I_{AE} < 0.1$ with an average error rate $\langle P_e \rangle \simeq 0.393$ for a combined application of two basis transformations.
\par 
These results have a direct impact on the security of such protocols. Due to the reduced information of an adversary and the high error probability introduced during the attack strategy, Alice and Bob are able to accept higher error thresholds compared to standard entanglement-based QKD protocols.

\bibliography{references}

\end{document}